\documentclass[9pt,conference]{IEEEtran}
\IEEEoverridecommandlockouts

\usepackage{amsmath,graphicx} 
\usepackage[hyphens]{url}
\usepackage{array, booktabs}

\newcommand{\ie}{\textit{i.\,e.,} }
\newcommand{\eg}{\textit{e.\,g.,} }

\title{AI-Generated Music Detection and its Challenges}

\author{
\IEEEauthorblockN{Darius Afchar}
\IEEEauthorblockA{
\textit{Deezer Research}
}
\and
\IEEEauthorblockN{Gabriel Meseguer-Brocal}
\IEEEauthorblockA{
\textit{Deezer Research}\\
{Paris, France}\\
{research@deezer.com}
}
\and
\IEEEauthorblockN{Romain Hennequin}
\IEEEauthorblockA{
\textit{Deezer Research}
}
}

\begin{document}

%
\maketitle
\begin{abstract}
In the face of a new era of generative models, the detection of artificially generated content has become a matter of utmost importance. In particular, the ability to create credible minute-long synthetic music in a few seconds on user-friendly platforms poses a real threat of fraud on streaming services and unfair competition to human artists. This paper demonstrates the possibility (and surprising ease) of training classifiers on datasets comprising real audio and artificial reconstructions, achieving a convincing accuracy of 99.8\%. To our knowledge, this marks the first publication of a AI-music detector, a tool that will help in the regulation of synthetic media.
Nevertheless, informed by decades of literature on forgery detection in other fields, we stress that getting a good test score is not the end of the story.
We expose and discuss several facets that could be problematic with such a deployed detector: robustness to audio manipulation, generalisation to unseen models.
This second part acts as a position for future research steps in the field and a caveat to a flourishing market of artificial content checkers.
\end{abstract}

\section{Introduction}

Generative models have gained tremendous popularity in the past couple of years.
Many discussions have ensued around the new opportunities these models may provide, as well as critics about their sociotechnical context and the many risks they entail. It is tricky to talk about generative AI in a neutral manner, mainly because this trending topic has a larger reach than purely technical considerations (\eg commercial, legal, social and political ramifications \cite{widder2024watchinggenerativeaihype}), and is fairly new for everyone. We also stress that so-called ``neutral'' discussions often support dominant views instead of truly enabling a scientific discourse encompassing all impacted stakeholders \cite{green2021data, crawford2021atlas}.
As motivation for this work, we therefore explicitly call for better regulation of these models, for a number of reasons that interested readers may find discussed in detail in \cite{goetze2024ai, gautam2024melting, gebru2023artists}. This constitutes a \textit{working assumption} that we will not further debate in this paper.

One of the many areas that needs to be addressed in regulating AI-music is to better identify synthetic generations within a body of genuine human-made content.
In 2023, a new wave of generative models has rendered the risks of \textit{AI-generated music} more tangible than before \cite{schneider2023mousai, agostinelli2023musiclm, garcia2023vampnet, copet2024simple}. Several user-friendly services have also recently emerged and democratised the creation and diffusion of AI-music 
: \eg Riffusion\footnote{\texttt{\url{www.riffusion.com}}}, Suno\footnote{\texttt{\url{www.suno.ai}}}, Udio \footnote{\texttt{\url{www.udio.com}}}, Stable Audio\footnote{\texttt{\url{stability.ai/stable-audio}}}.
AI-music now pose a growing problem for music artists and labels. Lawsuits are currently being filed against several AI companies \cite{widder2024watchinggenerativeaihype}.

While studies have been conducted on detecting synthetic sounds and singing voices \cite{wu2017asvspoof, zang2024singfake}, we present a novel setting in this paper. We propose the first general-purpose AI-generated music detector, a significant advancement that also includes generated instrumental parts. Our focus is on the trending waveform generators mentioned earlier. We leave symbolic or MIDI-based synthesis models for future exploration.
Using basic convolutional models, we show that almost perfect detection scores ($>$99\% accuracy) are easy to obtain.

Although AI-music detection is novel, we do not conduct our research in a vacuum.
This task is very reminiscent of the topic of artificial forgery detection, within the field of media forensics: \eg deepfake detection, image tampering, voice spoofing \cite{rossler2019faceforensics, wu2017asvspoof}. While these detectors are not directly transferable to the specifics of music, we can at least anticipate having to deal with similar research questions raised in this literature \cite{mirsky2021creation, lin2024detecting}. Therefore, in the second part of our paper, we take a step back on our seemingly impressive results and discuss caveats to AI detectors: robustness to audio manipulation and generalisation to unseen generators. In a nutshell, \textit{we have to look beyond performance scores}, no matter how good they look.

This paper serves as a first research study on AI-music detection and a proof of concept that they can be detected, but also as a position and message for the research community on the many facets and challenges to consider for the future research steps of this topic.

Our code is available at \texttt{\url{github.com/deezer/deepfake-detector}}.


\section{AI-music detection}

There are many ways to tackle AI-music detection. In this section, we first discuss our choice of framework and its advantage to solve the task, the employed data as well as some first surprisingly convincing detection scores.

\subsection{Proposed framework}

Motivated by the democratisation of online platforms that can generate minutes-long synthetic music, we restrict the scope of our paper to waveform based generators, common in these services: \eg 
MelGAN \cite{kumar2019melgan},
HiFiGAN \cite{kong2020hifi}, Jukebox \cite{dhariwal2020jukebox}, Musika! \cite{pasini2022musika}, Moûsai \cite{schneider2023mousai}, MusicLM \cite{agostinelli2023musiclm},
VampNet \cite{garcia2023vampnet}, MusicGen \cite{copet2024simple}.
This list is not exhaustive of the swarm of published music generation methods. We only provide a representative subset.

While it is impossible to account for all particularities, we can usually break down these models into two parts. First, an autoencoder (AE) is trained to compress bits of raw audio into an easier representation to process and to invert this representation into an audio signal (\ie vocoder).
For instance, mel-spectrogram representations were often used (\eg \cite{kumar2019melgan}) before being replaced by more recent so-called neural codecs -- as Soundstream \cite{zeghidour2021soundstream}, Encodec \cite{defossez2022high} or DAC \cite{kumar2024high} -- that demonstrated better reconstructions. For interested readers, the latter commonly employ discretised latent spaces as codebooks of tokens -- \eg \textit{Residual Vector Quantization} (RVQ) \cite{razavi2019generating}. Then, a second internal module is usually trained to learn to continue the compressed sequence temporally or generate it conditioned on text inputs, depending on the considered task. For instance, large-language-model (LLM) inspired architectures have been proposed for the role \cite{garcia2023vampnet}.
Put simply, the AE does the waveform synthesis part while the LLM does the semantic work of generating a coherent musical sequence through time.

Detecting that a music sequence was artificially generated can be tricky. With the risk of falling into anthropomorphism, this equates to trying to learn a musician's style: \eg MusicGen might always generate music with a specific musical structure. Conversely, it might be easier to try to catch if an audio sample is the output of an AE. For instance, it is well-known that neural decoders tend to produce \textit{checkerboard artefacts} \cite{kumar2019melgan} characteristic of transposed convolution operations. We might be able to catch many more such artefacts. Thus, we propose the following research direction: \textbf{can we detect if a music sample is generated by an artificial decoder, this independently of its musical content?}

Another difficulty is of a causal nature. If we collected real and synthetic music samples and naively trained a model to classify them, we might end up detecting features unrelated to generation artefacts. For instance, a public real music dataset might be full of classical music, while a synthetic dataset primarily includes rap and pop music. This could result in the classifier learning to detect classical music instead of distinguishing real and forgeries. This problem is known as \textit{confounding} \cite{peters2017elements}.
The same discussion applies to the compression codec that might confound the detection of AI-music (\eg all Riffusion songs are exported in mp3 192kbps).

These two remarks have motivated our following framework: We leverage a dataset of real music samples, which we auto-encode using the trained AE part of the above models. These samples are stored at the same bitrate as the original audio. Controlling on the music semantic and file encoding, the model we then train can only detect generation artefact since it should learn to tell apart a real audio from its reconstructed counterpart. Therefore, we limit these extraneous confounding influence \cite{peters2017elements}, \ie shortcut learning.

\subsection{Considered dataset and music generators}

We chose to use the FMA dataset \cite{fma_dataset}, an open dataset that allows reproducibility and comparison of future work. Due to size constraints, we only consider the medium split that includes 25.000 music tracks spread into 16 genres. All tracks are encoded in mp3 with a diversity of bitrates -- with a majority of 320kbps, followed by 256 and 192kbps. The audio files are processed in 44.1kHz.

As for the autoencoders, we consider two popular neural codecs: Encodec (\eg used in \textit{Suno's Bark} and MusicGen) and DAC (\eg used in Vampnet\footnote{For interested readers, we actually use the LAC version of DAC that is better suited for music, similar to what is done in VampNet \cite{garcia2023vampnet}.}). We also studied the decoder part of Musika, which was trained end-to-end on polar spectrograms. Finally, we consider a combination of a mel-spectrogram converter-inverter and a Griffin-Lim phase reconstruction (\eg used in \textit{Riffusion v1}) -- we dub this pipeline \textit{GriffinMel}. Some audio reconstructions are available on our repository to gain intuition on each decoder.
The availability of trained models constrained our choice of decoder. For instance, Soundstream is used as a latent representation in many of Google's models, yet no public checkpoint is available. The same applies to MelGAN and HiFiGan, for which no checkpoint exists for music data\footnote{Although there exists some for voice synthesis, we found that this resulted in heavy audible artefact on music that we deemed unrealistic.}, as well as \textit{Riffusion v3} and \textit{Suno's Chirp} that are now closed-source.
We consider several configurations for the above decoders: Encodec in 3, 6 and 24kbps, DAC in 2, 7 and 14kbps, and GriffinMel using 256 and 512 melbands.

We autoencode all considered real tracks and obtain nine different autoencoded reconstructions: \ie with the same ``semantic'' musical content and stored with the same bitrate, but with different artefacts linked to each AE. This leads to a dataset of 250.000 tracks.
We split the songs (and their corresponding reconstructions) in a 70\%, 10\%, 20\% fashion between train, validation, and test.
Empirically, the GriffinMel reconstruction seems the easiest to catch due to audible phase errors.
Encodec and DAC sound most challenging, especially at their maximum quality setting. If it is often possible to distinguish between a real and a reconstruction when placing one next to the other, it is way more tricky without a point of reference or being aware that the audio could be generated. This relates to the recent user study in \cite{cooke2024good}.

\begin{table}[t]
\centering
\caption{Test detection scores for different audio representations.}
  \begin{tabular}{cc}
    \toprule
    Model&Accuracy (\%)\\
    \midrule
    waveform & 95.2 \\
    complex & 92.9 \\
    \textbf{amplitude} & \textbf{99.8} \\
    phase & 99.6 \\
    polar & 99.7 \\
  \bottomrule
\end{tabular}
    \label{table:results}
\end{table}

\subsection{First results of detection}

We started experimenting with our dataset with straightforward convolutional models (\ie alternating convolution and pooling layers). To our surprise, this basic setting led to test accuracies over 90\%, which we were initially sceptical about. After several experiments, it seems this detection task is easier than we thought. As we will see next, our high scores did not prompt us to explore more complex models but rather to sanity check an already good performing model.

Briefly, our proposed model is composed of six convolutional layers with $[16, 32, 64, 128, 256, 512]$ filters. We use kernels of size 3 and a pooling of size 2. We finish with an average pooling and two linear layers.
During training, real and synthetic music tracks are sampled with a $\frac 1 2$ probability. The synthetic one is sampled uniformly among the nine reconstructions.
We extract a random 0.8s snippet from each track of the batch\footnote{Fixed arbitrarily and leading to 128 STFT time frames.}.
We use some common data preprocessing and augmentation of the audio: STFT, random mono mix, random gain, frequency cutoff at 16kHz\footnote{... to avoid relying on spurious mp3 conversion artefacts.}, and conversion to decibel scale when applicable.
All details and trained weights are available on our repository.


\begin{table*}[h]
    \centering
    \caption{{Robustness accuracy test scores.} 
    \normalfont{
    We test the accuracy of the amplitude-spectrogram-based model on various audio transforms (\%).
    We include a breakdown per class.
    We report the previous \textit{amplitude} model scores for comparison on top.
    Background colours highlight \textit{score degradation}.}
    }
    \includegraphics[width=\linewidth]{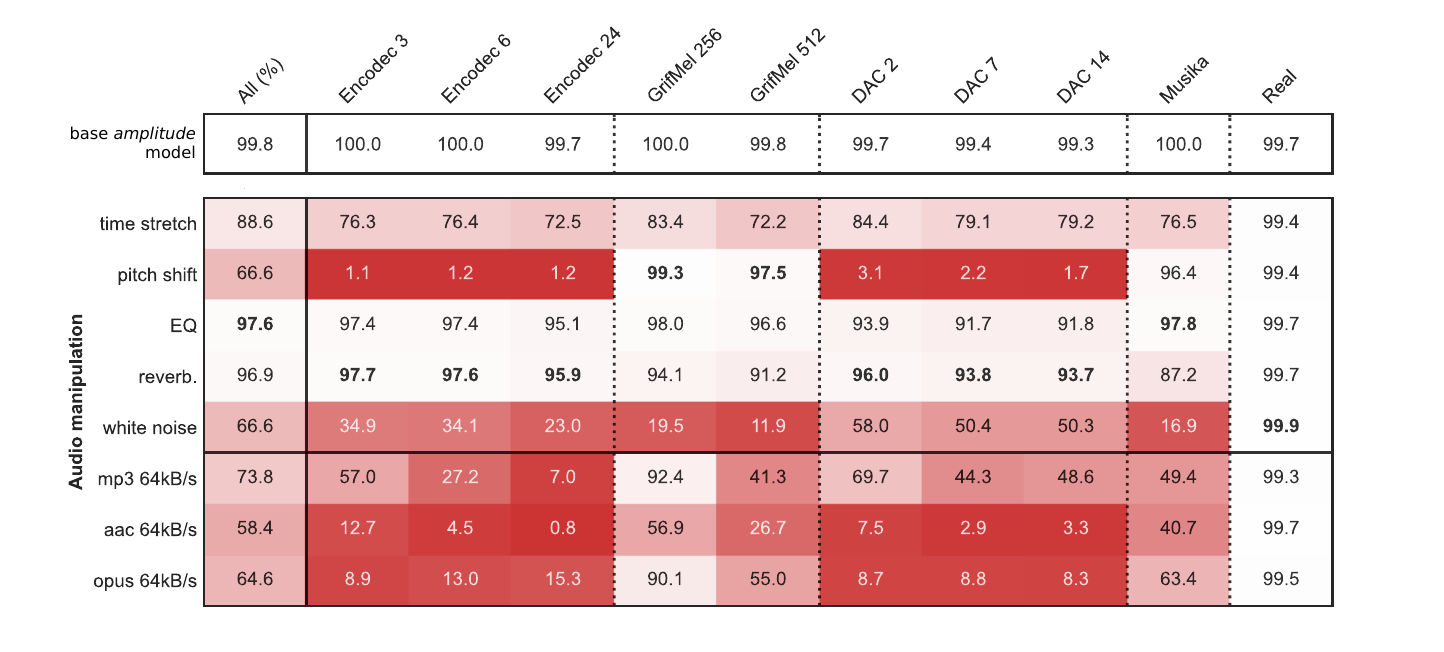}
    \label{table:robustness}
\end{table*}

The architectural choices does not seem to impact the performance much. However, the choice of input preprocessing seems to be much more influential on our experiment results. We report in Table \ref{table:results} the detection score for various choices of audio representations: the raw waveform, the complex STFT, its amplitude, its phase, or both stacked as polar coordinates. All our results are very satisfying overall.
Transforming music samples as \textit{amplitude} spectrograms leads to the best performance overall (\textbf{99.8\%} accuracy).
It is also interesting to see that the purely \textit{phase}-based model yields high scores despite often being considered less efficient than the amplitude representation.
A per-class breakdown is available in Table \ref{table:robustness} and highlight consistent performances across different AE.

\subsection{Generalisation to full music generators}

We argued that instead of learning to detect AI-music generators, we could more simply learn to detect the fingerprint of the AE they employ. As a sanity check, we test our model on synthetic music created from text prompts, \ie unseen during training, not autoencoded nor related to FMA. We gather 50 tracks from MusicGen, amounting to 25 minutes of music, and extract random snippets from them. On a total of 2500 music snippets, we achieve a \textbf{99.9\%} detection score.

\vspace{0.5em}
We underscore that given our resulting performances, we did not find it so crucial to explore the best possible architecture further and deemed it more important to discuss the aftermath of obtaining such convincing scores. Indeed, it could feel that we have ``solved'' the task. Nevertheless, should we be so confident?

\section{Caveats on AI-music detectors}

Beyond the lab experiments, we find it crucial to ponder the consequence of deploying AI-music detectors to the world.
Indeed, a new market of ``AI content detectors'' has emerged in recent years: \eg checking if a student essay employed ChatGPT\footnote{\eg \texttt{\url{gptzero.me}}}.
Such tools often claim to have high detection scores. However, they are often closed-source, making verification tricky.
This has notably led to strange situations where students have much trouble proving their good faith in false positive cases against the ethos of a so-called ``trusted AI checker'' \cite{url_ai_detection}.
Several commercial solutions have also recently been released for AI-music\footnote{\eg \texttt{\url{ircamamplify.io}},\, \texttt{\url{pex.com}}}. So far, they have followed the same path as AI-text detectors: closed-source and without any associated research publication. This does not allow rigorous studies~\cite{casper2024black}.

This section hence discusses aspects to make AI-music detection more realistic and reliable.
We discuss two main facets that make the detection more complex than may first appear: its robustness to audio manipulation and its generalisation to unknown encoders.
We also highlight how untrustworthy performance numbers can be, which calls to make these detectors open source and considering other aspects to validate a model (\eg interpretability).

\begin{table}[h]
    \centering
    \caption{{Generalisation to unseen music generators.} \normalfont{We train on each single decoder indicated on the left axis and evaluate on each test accuracy (\%) of the bottom axis.}
    }
    \includegraphics[width=\linewidth]{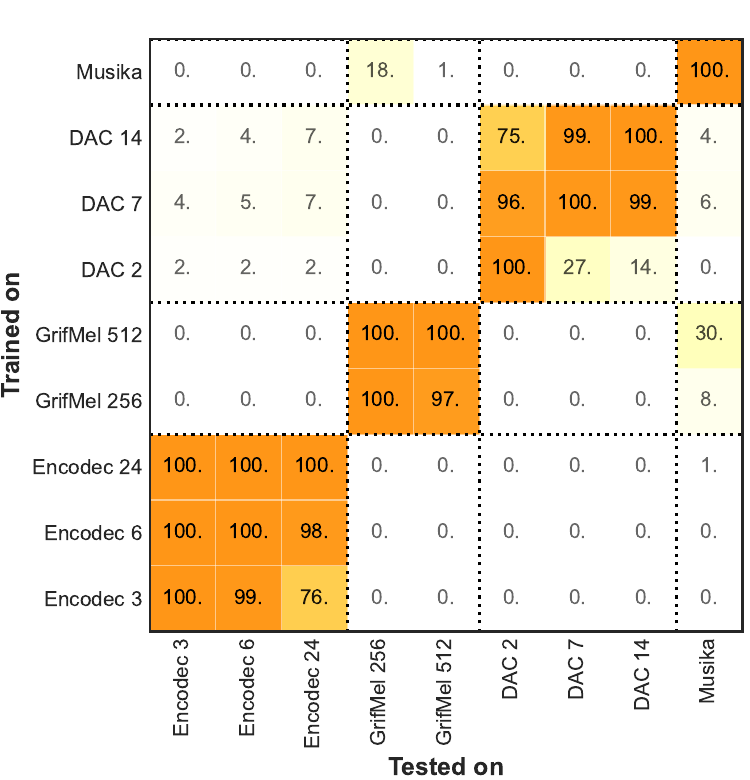}
    \label{table:generalisation}
\end{table}

\subsection{Robustness to manipulations}
\label{sec:robustness}

An angle often discussed in the literature on forgery detection is the robustness to data shifts.
There are countless scenarios where AI-music creators do not directly publish the immediate output of the generative model. For instance, they could genuinely reencode them in a different format while exporting the result or adding it to a video clip. They could also try to bypass a detector more strategically by applying time-stretching or pitch shift transforms, similar to what is frequently done on social networks to bypass fingerprint systems and evade copyright claims.
It would be unrealistic not to expect some users to try to evade detection.

As a first study, we consider some common audio transformations that lay users could employ: random pitch shift ($\pm2$ semitones), time stretch ($[80,120]\%$), EQ, reverb, addition of white noise, reencoding in \textit{mp3}, \textit{aac}, and \textit{opus} in 64kbps.
Implementation details are available in our repository.
We leave attacks from more advanced users for future work (\eg adversarial attacks \cite{casper2024black}).

We evaluate the amplitude-based model from the previous section on such unseen transformation and report the results in Table \ref{table:robustness}.
The performances drop drastically under pitch shifts, the addition of white noise, and codec reencoding. This is consistent with previous literature on forgery detection that ML models are generally not robust to out-of-distribution shifts -- if not explicitly designed for them \cite{wang2020cnn, li2020celeb}.
Conversely, it is unclear why the model remains robust to some manipulations (\ie time stretch, EQ and reverberation).

We highlight that several scores drop to almost zero, which means that the model has predicted the real class for most samples (instead of more unconfident, aleatoric guesses). Meanwhile, the manipulations did not impact the \textit{real} class scores. 
This suggests that the model works by detecting artefacts specific to each AE and otherwise defaulting to the \textit{real} class if none is found (or that the manipulations make them unrecognisable for the model). This would not be surprising since the \textit{real} class can be expected to be more diverse and complex than autoencoded generations \cite{tishby2000information}. Since ML models are biased toward simple solutions \cite{scimeca2021shortcut}, it is expected that it is easier to detect a real audio by \textit{\underline{not} detecting the characteristics of generated samples} instead of learning a manifold of real music.
This is a critical remark because we can already anticipate that the model may not generalise to unseen music generators.

\subsection{Encoder generalisation}

Another important question is whether our detector generalises to AE models that were not considered during training. Instead of finding additional AE to test, we conduct a new experiment in which we retrain our best model from scratch on each of the nine considered decoders (versus real audios) and check how the detection performance naturally transfers to the others. The results are displayed in Table \ref{table:generalisation}.
Interestingly, we first find that the models are pretty robust \textit{intra}-family: \eg learning on Encodec 24kbps reconstruction transfers well to 6kbps and 3kbps. It is reassuring that we may not need to include all possible parametrisation of an AE to learn to detect it.
Learning from a higher bitrate seems to transfer better to low bitrate, which could stem from the RVQ formulation of the considered models, but this is not so straightforward to assert.
Then, we note that the model falters on \textit{inter}-family generalisation: said performances are almost always zero (\eg GriffinMel $\rightarrow$ DAC).
This aligns with the previous section that the models are not robust to unseen manipulations.
Note that the performances drop again to 0\%, which implies that the \textit{real} class may be acting as a default.

\subsection{Challenges ahead}

We did not train the models on the audio manipulations of Sec.~\ref{sec:robustness} (\ie data augmentation). We studied their natural robustness.
In some subsequent experiments, we saw that fine-tuning on these manipulations could reliably restore high accuracy scores. The same is true about fine-tuning to a new decoder. 
However, in this paper, we prefer to insist on the following: \textit{there will always be an unseen manipulation or generation method}. It would not be realistic to only evaluate our model on data and settings we optimise for.

In the long run, this is a cat-and-mouse game, where it is illusory to anticipate all cases in advance. In particular, attackers will always find new ways to evade detection, and new models will be released \cite{mirsky2021creation, lin2024detecting}.
Overall, our results suggest that straightforward AI-music detectors are not naturally robust to such unanticipated cases.
We believe this calls for a much more continual process of patching a detector regularly (\ie similar to an antivirus software).
Evaluating detectors in a scenario of partial knowledge is also essential to reveal their limits and how they handle unusual inputs.

Instead of solely focusing on accuracy, our experiments may call to working on making AI-detectors more \textit{interpretable}, thus enabling to debug the sanity of a prediction (\eg to handle false positives).
The phenomenon we uncover that the \textit{real} class acts as default also exposes that the probabilities that our model output should not be taken at face value as a ``percentage of AI content''. This relates the topic of model \textit{calibration} (\eg how detectors should be calibrated to handle audios mixing real and synthetic stems) and more largely \textit{specification} on how a system is expected to function.

Lastly, let us acknowledge that regulating AI-music with AI-detector is a form of techno-solutionism.
It can lead to a myopic view of the topic, potentially overlooking other parts of the full AI supply chain \cite{parliament2021, miotti2024combatting}. For instance, it might be more efficient to regulate big tech actors, than putting off fires of detecting these generations afterwards. An alternative lead could be to have them employ watermarking, preventing the bulk of lay users from spreading unlicensed generations. However, this technique is far from flawless \cite{barman2024brittleness}.

\section{Conclusion}

In this paper, we proposed the first study on AI-generated music detection. We show that such forged content is surprisingly easy to detect, yet stress that a good accuracy score is not at all the end of the story and recommend considering several additional aspects (\eg robustness to manipulation, generalisation to unseen settings).


Our future work includes studying whether these models can be easily fine-tuned or updated for new generators, their generalisation capabilities with further data augmentation during training (\eg audio manipulations), defense against adversarial attacks, interpretability, and the impact of more realistic stem mixing and audio engineering.


\bibliographystyle{IEEEbib}
\bibliography{strings,refs}

\end{document}